# Revealing curvature and stochastic effects on grain growth kinetics: A thermodynamic perspective from extremal principle

*Yue Li, Zhijun Wang\*, Junjie Li, Jincheng Wang*

*State Key Laboratory of Solidification Processing,*

*Northwestern Polytechnical University, Xi'an 710072, China*

**Abstract:**

In the theoretical development of normal grain growth, the roles of 'drift' (curvature effect) and 'diffusion' (stochastic effect) have been an open question for many years. By coupling contributions of microstructure entropy and grain topological interactions with thermodynamic extremal principle (TEP), this letter extends the existing thermodynamic framework of grain growth. It not only explains the curvature and stochastic effects by thermodynamics but provides a new mean-field model for grain growth. Through thermodynamic modification, this new model can yield a high-accuracy representation of existing ultra-large phase-field simulated results.

**Keywords:** Grain growth, Thermodynamics, Kinetics, Extremal principle

As one of the most fundamental phenomena in microstructure evolution, grain growth is a theoretical subject of long-standing interest in the context of physical metallurgy and other scientific disciplines. Despite numerous studies in the past 70 years, the true picture of its physics is still away from complete, e.g., the dominant factors of scaling behaviors in grain growth. Up to now, the evolution of grain size distribution has been considered driven by 'drift' (curvature effect) solely **[1-7]** or co-controlled by 'drift' and 'diffusion' (stochastic effect) together **[10, 11, 13-16]**, which have induced many discussions of different viewpoints like Refs. **[12, 16]**.

---
\* Corresponding author. zhjwang@nwpu.edu.cn

The 'drift' theory of grain growth was initiated by Hillert [1], who assumed the change rate of an individual grain as

$$\frac{dR}{dt} = \alpha M\sigma\left(\frac{1}{R_{cr}} - \frac{1}{R}\right), \qquad (1)$$

where $\alpha$ is a geometrical constant, $M$ and $\sigma$ denote the grain boundary mobility and energy, respectively. $1/R$ is the curvature of individual grain, and $1/R_{cr}$ is the mean curvature of the total system. Hence, the curvature difference determines the grain boundary migration and the flux in radius space $J$, and continuity equation for grain size distribution $F(R,t)$ yields as

$$\frac{\partial F(R,t)}{\partial t} + \frac{\partial J}{\partial R} = 0; \ J = J_{Dr} = \frac{dR}{dt}F(R,t), \qquad (2)$$

where $J_{Dr}$ is the 'drift' flux in radius space. Afterward, in consideration of the spatial correlation, Hundari, Ryum, and their colleagues [2-4] modified Eq. (1) by

$$\frac{dR}{dt} = \alpha M\sigma\left(1 + \beta_1 \frac{R}{R_{cr}} + \beta_2 \left(\frac{R}{R_{cr}}\right)^2 + ...\right)\left(\frac{1}{R_{cr}} - \frac{1}{R}\right), \qquad (3)$$

which is analogous to Streitenberger and Zöllner's effective growth law [5, 6], whose quadratic approximation can be thought as a first-order approximation of Eq. (3). Recently, applying the same framework in the 3-D grain growth [7], the authors of this paper found that almost all necessary statistical topological relationships [8, 9] of grain growth can be consistently derived from the first-order approximation of Eq. (3), showing the validity of the pure 'drift' theory for normal grain growth behaviors.

Besides the former pure 'drift' theories, researchers have proposed two different models of combing the curvature and stochastic effects together. It means that the flux in radius space consists of 'drift' flux and 'diffusion' flux simultaneously. Combining Hillert's mean-field theory [1] and Mulheran and Harding's stochastic theory [10],

Helfen et al. **[11]** gave a Fokker–Planck type equation for describing evolutions of grain size distributions

$$\frac{\partial F(R,t)}{\partial t} + \frac{\partial}{\partial R}[(J_{Dr} + J_{Di})] = 0;$$
$$J_{Dr} = \alpha M\sigma\left(\frac{1}{R_{cr}} - \frac{1}{R}\right)F(R,t);\ J_{Di} = -D\frac{\partial}{\partial R}\left(\frac{1}{R^d}\frac{\partial F(R,t)}{\partial t}\right)F(R,t), \quad (4)$$

where $J_{Di}$ denotes the 'diffusion' flux in radius space, $D$ is the 'diffusion' coefficient, and $d$ equals 1 and 2 for the 2-D and 3-D cases, respectively. They interpreted their stochastic term by the atomic jumps along the grain boundary and found it only makes sense for the nanocrystalline system. Such viewpoints are consistent with Mullins's discussion **[12]**. In parallel, based on Loudt's stochastic work **[13]**, Pande and his cooperators also developed a combined model **[14-16]**, whose Fokker–Planck type equation was represented by

$$\frac{\partial F(R,t)}{\partial t} + \frac{\partial}{\partial R}[(J_{Dr} + J_{Di})] = 0;$$
$$J_{Dr} = \left(\frac{b}{R_{cr}} - \frac{a}{R}\right)F(R,t);\ J_{Di} = -D\frac{\partial}{\partial R}\left(\frac{\partial F(R,t)}{\partial t}\right)F(R,t), \quad (5)$$

where $a$ and $b$ are two constants. This stochastic term was explained by Pande with grain-environmental heterogeneity **[1]**. And Eq. (5) can predict an exact same grain size distribution function as the first-order approximation of Eq. (3), fitting well with simulations and experiments in recent years **[7]**.

However, both stochastic approaches are based on a prior assumption of a Fokker–Planck (FP) equation, whose validity for microstructure evolution is not proved yet. For example, like Eq. (3) of grain growth, the classical mean-field treatment **[17]** can also yield an exact prediction of Ostwald ripening kinetics without assuming any FP equation. This motivates us to verify the existence of the Fokker–Planck equation from a thermodynamics approach, i.e., the thermodynamic extremal principle (TEP) **[18]**.

As one of the well-approved approaches of microstructure evolution, the thermodynamic extremal principle has also been applied in grain growth, where a pure 'drift' equation Eq. (1) was reproduced by Svoboda and Fischer **[18, 19]**. However, both spatial correlation [Eq. (3)] and the stochastic effect [Eqs. (4) and (5)] are negated by the existing TEP-based model, which may be attributed to an overlook of some potential thermodynamic characteristics in grain growth.

As a first step, we account for the contribution of entropy. The entropy of grain growth structure, defined as the microstructure entropy, was first proposed about ten years ago **[20]**. Recently, this concept began to attract more attention because of its relevance to the material's properties **[21, 22]**. To couple such microstructure entropy into the total free energy $G$, Svoboda and Fischer's distribution concept **[18, 19]** is adopted here

$$G = \int_0^\infty \{F(R,t)\left[2\pi R^2 \sigma + k_B T \ln(F(R,t))\right]\}dR, \qquad (6)$$

where $-k_B \cdot F(R,t)\ln(F(R,t))$ represents the microstructure entropy, $k_B$ is the Boltzmann constant, $T$ is the temperature, and $2\pi R^2 \sigma$ denotes the energy of an individual grain by spherical approximation. The upper limit of this integral is chosen to be infinite $\infty$, despite the existence of the maximum radius $R_{\max}$. It means the distribution probability of the interval $R_{\max} \to \infty$ will naturally go to 0, as a piecewise function in mathematics. With the fixed radius space, the state of the system (total free energy) will be uniquely determined by $F(R,t)$. Thus, the change rate of Eq. (6), $\dot{G}$, yields as

$$\dot{G} = \int_0^\infty \left\{\frac{\delta \tilde{F}}{\delta F(R,t)}\frac{\partial F(R,t)}{\partial t}\right\}dR = \int_0^\infty \left\{\frac{\delta \tilde{F}}{\delta F(R,t)}\left[\frac{\partial}{\partial R}(-J)\right]\right\}dR, \qquad (7)$$

where $J$ still represents the total flux in radius space. In TEP **[16]**, the dissipation functional $Q$ follows a quadratic form

$$Q = \int_0^\infty \frac{J^2}{M_J} dR, \tag{8}$$

where $M_J$ denotes the general dissipation mobility. After applying the constraint for dissipation functional, $\dot{G} + Q = 0$, the thermodynamic variation yields as **[23]**

$$\frac{\delta}{\delta J}\left\{\dot{G} + \frac{Q}{2} + \lambda\left[V - \int_0^\infty \frac{4\pi R^3}{3} F(R,t) dR\right]\right\} = 0, \tag{9}$$

where $V$ is the total volume and $\lambda$ is the Lagrange multiplier for the constraint of the volume conservation. Therefore, the total flux in radius space yields as

$$J = -M_J\left\{4\pi R\sigma - \lambda 4\pi R^2 + \frac{k_B T}{F(R,t)}\frac{\partial F(R,t)}{\partial R}\right\}. \tag{10}$$

In Svoboda and Fischer's dissipation of distribution concept **[18, 19]**, the total dissipation is proportionate to the $F(R,t)R^2/M$. Similarly, we here denote the general dissipation mobility $M_J$ as

$$M_J = \frac{\alpha F(R,t)}{4\pi R^2} M = \frac{\alpha F(R,t)}{4\pi R^2}\frac{D_0}{k_B T}, \tag{11}$$

where $D_0$ is the diffusion constant of atoms and $M = D_0/k_B T$ is the famous Einstein-Nernst relation. Now, Eq. (10) can be represented by

$$J = -\left\{\alpha M\left(\frac{\sigma}{R} - \lambda\right) F(R,t) + \frac{\alpha D_0}{4\pi R^2}\frac{\partial F(R,t)}{\partial R}\right\}. \tag{12}$$

Taking it into the continuity equation, a Fokker–Planck type equation for grain growth can be deduced

$$\frac{\partial F(R,t)}{\partial t} = \frac{\partial}{\partial R}(-J)$$

$$\frac{\partial F(R,t)}{\partial t} = \frac{\partial}{\partial R}\left\{\left[\alpha M\left(\frac{\sigma}{R} - \lambda\right)\right]F(R,t) + \frac{\partial}{\partial R}\left[\frac{\alpha D_0}{4\pi R^2}\frac{\partial F(R,t)}{\partial R}\right]\right\}. \tag{13}$$

With the constraint of total volume ($dV/dt = 0$), we have

$$\frac{dV}{dt} = \int_0^\infty \frac{4\pi R^3}{3} \frac{\partial F(R,t)}{\partial t} dR = \int_0^\infty \frac{4\pi R^3}{3} \frac{\partial}{\partial R}\left\{\left[\alpha\left(\frac{\sigma}{R} - \lambda\right)\right] F(R,t)\right\} dR \\ + \int_0^\infty \frac{4\pi R^3}{3} \frac{\partial}{\partial R}\left[\frac{\alpha D_0}{4\pi R^2} \frac{\partial F(R,t)}{\partial R}\right] dR = 0. \quad (14)$$

Since $F(R,t)$ and $\partial F(R,t)/\partial t$ approach to 0 when $R \to 0$ and $R \to \infty$, after integral by part, Eq. (14) becomes

$$\int_0^\infty 4\pi R^2 F(R,t) dR = -\int_0^\infty 4\pi R^2 \left[\alpha M\left(\frac{\sigma}{R} - \lambda\right)\right] F(R,t) dR, \quad (15)$$

and the multiplier $\lambda$ yields as

$$\lambda = \frac{\int_0^\infty R F(R,t) dR}{\int_0^\infty R^2 F(R,t) dR}, \quad (16)$$

which is the same as that of Fischer and Svoboda's works **[18, 19]**. Defining the $\lambda = 1/R_{cr}$, Eq. (13) will transform into

$$\frac{\partial F(R,t)}{\partial t} = \frac{\partial}{\partial R}\left\{\left[\alpha M\left(\frac{\sigma}{R} - \frac{\sigma}{R_{cr}}\right)\right] F(R,t) + \frac{\partial}{\partial R}\left[\frac{\alpha D_0}{4\pi R^2} \frac{\partial F(R,t)}{\partial R}\right]\right\}. \quad (17)$$

Taking $D = \alpha D_0/4\pi R^2$, this equation is exactly consistent with Helfen et al.'s hypothesis of Eq. (5) **[9]**. This consistency indicates that the stochastic effect originates from the microstructure entropy and works only for the nanocrystalline system. In other words, despite the potential influences on material's properties **[21, 22]**, the microstructure entropy won't significantly affect the microstructure evolution kinetics.

However, when the stochastic term of Eq. (17) is neglected, the remaining 'drift' equation will coincide with Hillert's model **[1]**, whose predictions of distribution will differ from the reality **[7]**. Recently, to order to interpret Kamachali et al.'s simulated $\alpha$ value and distributions **[24, 25]**, Svoboda and Fischer have added an extra translational dissipation in their TEP framework **[26]**. It provides us with a promising direction of explaining $\alpha$'s value but limited by the predicted bimodal-like grain size distribution. Thus, translational dissipation is not employed in this work for the explanation of real grain size distribution. In contrast, in consideration of the well simulated and observed grain topological correlation **[7]**, the grain contact affinity is believed to determine the real grain growth kinetics.

Grain contact affinity has been found to be closely related to the curvature difference, where grains with higher curvatures of opposite signs will exhibit higher contact affinities **[27]**. Recently, Lutz et al. **[28]** and Lazar et al. **[29]** defined a purely topological energy to establish thermodynamic-like theories of describing grain topology behaviors. Inspired by these heuristic works of grain topology-curvature correlation, aimed at reflecting the grain contact affinity, the present TEP framework [Eq. (6)] should also contain a curvature-associated energy.

Since the relative radius $\rho = R/R_{cr}$ is an indicator of the curvature difference, we here express the curvature-associated energy with a phenomenological polynomial of $\rho$

$$G = \int_0^\infty F(R,t)\left[2\pi R^2 \sigma + k_B T \ln(F(R,t)) + P(\rho)\right]dR; \quad (18)$$
$$P(\rho) = 2\pi R^2 \sigma (C_1 \rho + C_2 \rho^2 + ...),$$

where $2\pi R^2 \sigma$ ahead of the polynomial is for the requirement of dimensions and $C_i$ are phenomenological parameters. Different from Lutz et al.'s **[28]** and Lazar et al.'s **[29]** theories, the topological energy defined here is desired to reflect the interactions between different types of grains (like the classical solid solution theory), rather than energies of individual ones. This integral should not exceed that of Eq. (6) since the whole grain boundary is only determined by the total grain boundary area. In other words, the present addition of grain interaction energy is only to modify the oversimplified expression of grain energy. Therefore, the construction of polynomial should satisfy

$$\int_0^\infty F(R,t) 2\pi R^2 \sigma (C_1 \rho + C_2 \rho^2 + ...) dR = 0. \tag{19}$$

Now, we can reevaluate the previous modified mean-field model of Eq. (3) (pure 'drift theory'). It is easy to find the following expression of interaction term can yield the same equation as the first-order approximation of Eq. (3) (after neglecting the unnecessary stochastic effect)

$$\int_0^\infty F(R,t) 2\pi R^2 \sigma \left(\frac{\beta}{3}\rho - \frac{\beta}{4}\rho^2\right) dR = 0, \tag{20}$$

$$\frac{dR}{dt} = \alpha M \sigma \left(1 + \beta_1 \frac{R}{R_{cr}}\right)\left(\frac{1}{R_{cr}} - \frac{1}{R}\right). \tag{21}$$

With distribution function in **Ref. [7]**, this integral is close but not equal to 0, meaning that Eq. (21) does not strictly obey the thermodynamic constraints. However, if applying higher-order approximation in Eq. (3) to eliminate the excess part, there will be no solution of the size distribution (see **Supplementary Materials**), showing the deficiencies of existing mean-field theory **[2-4]**. To solve it, a new type of modification is adopted, whose construction of polynomial satisfies

$$\int_0^\infty F(R,t) 2\pi R^2 \sigma \left(\frac{\beta_1}{3}\rho - \frac{\beta_1}{4}\rho^2 + \frac{2\beta_2}{5}\rho^{\frac{1}{2}} - \frac{2\beta_2}{7}\rho^{\frac{3}{2}}\right) dR = 0, \tag{22}$$

and its corresponding equation of change rate of radius yields as

$$\frac{dR}{dt} = \alpha M \sigma \left[1 + \beta_1 \frac{R}{R_{cr}} + \beta_2 \left(\frac{R}{R_{cr}}\right)^{\frac{1}{2}}\right]\left(\frac{1}{R_{cr}} - \frac{1}{R}\right). \tag{23}$$

Although another tuning parameter is added here, the number of independent parameters does not change with the constraint of Eq. (22). Through the numerical iterations in **Supplementary Materials**, the solved grain size distribution family is shown in **Fig. 1**. The shape of distribution will change with different values of $\beta_1$, which represents the different grain topology correlations.

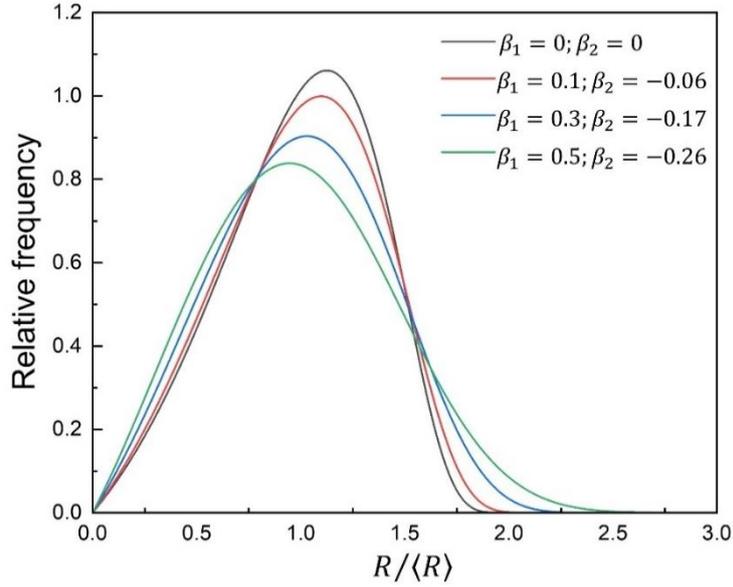

Fig. 1. The present grain size distribution family by Eqs. (21) and (23) under different $\beta_1$.

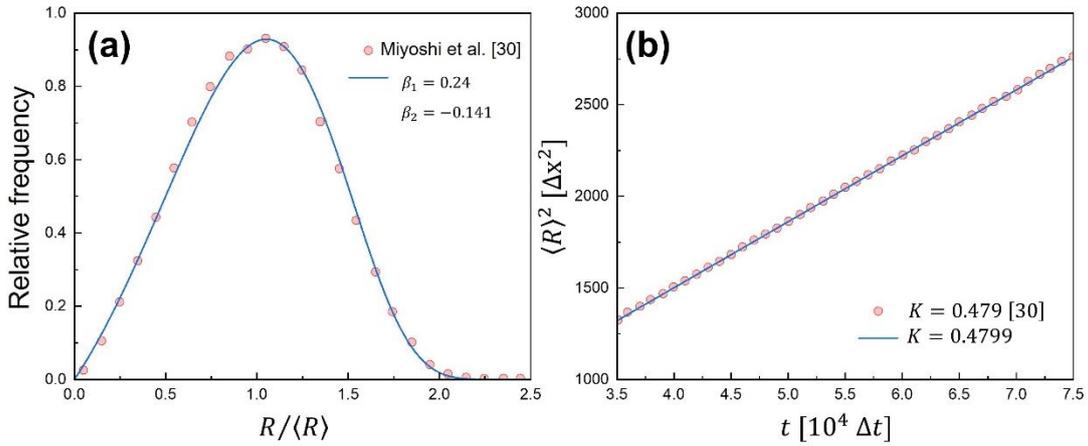

Fig. 2. Comparison of present (a) grain size distribution; and (b) growth rate of the squared average radius (Eqs. (21) and (23) with $\alpha=1$ **[1]**) with Miyoshi et al.'s phase-field simulation results **[30]**.

To compare with the recent ultra-large phase-field simulations, Kamachali et al.'s geometrical factor is adopted **[25]**, from which we obtain that $\beta_1 \approx 0.24$ and $\beta_2 \approx -\beta_1/1.7$ (see **Supplementary Materials**). **Fig. 2** exhibits the solved grain size distribution and average growth rate, which is a high-accuracy representation of Miyoshi et al.'s results **[30]**. It also proves the validity of the present TEP-based grain growth theory. Furthermore, because Eq. (5) and Eq. (21) can yield almost the same kinetic description of grain growth, Pande et al.'s stochastic theory **[14-16]** is only thought to be an alternative description of the grain-environment heterogeneity, analogous to the recent computational stochastic approach of simulating grain growth **[31]**. It is also consistent with Ng's statistical mechanical model of one-dimensional normal grain growth **[32]**, where there is no Loudt's diffusion term **[13]**.

In conclusion, this work reveals the previous controversy on curvature and stochastic effects in grain growth kinetics from a thermodynamic perspective. Originating from the microstructure entropy, the stochastic effect of atomic jumps on grain growth kinetics is revealed to make sense only for the nanocrystalline systems, consistent with the previous conclusions of Mullins and Helfen et al. Furthermore, inspired by recent works of grain topology correlation, a new mean-field model of grain growth can be derived by expanding the system's free energy by a polynomial of relative radius. Without increasing the number of independent variables, this model shows a high-accuracy representation of existing ultra-large phase-field simulation results. And Pande et al.'s systematic stochastic description of the grain-environment heterogeneity is the first-order approximation of the present model. In the future, it is of great significance to continue revealing the underlying thermodynamics of grain topology correlation.

**Acknowledgments**

The work was supported by the Research Fund of the State Key Laboratory of Solidification Processing (NPU), China (Grant No. 2020-TS-06, 2021-TS-02). Y. Li sincerely thanks for the attractive graduate course '*Thermodynamic Extremal Principle and Its Application*' by Prof. Haifeng Wang in the *Center of Advanced Lubrication and Seal Materials, Northwestern Polytechnical University*.

Supplementary Material for

# Revealing curvature and stochastic effects on grain growth kinetics: A thermodynamic perspective from extremal principle

*Yue Li, Zhijun Wang\*, Junjie Li, Jincheng Wang*

*State Key Laboratory of Solidification Processing,*

*Northwestern Polytechnical University, Xi'an 710072, China*

This Supplementary material is applied to help readers clearly see the calculation procedure and details of the grain growth kinetics and better understand the advantage of the present model in the main text.

The calculation of growth rate and size distribution are based on the famous Lifshitz- Slyozov- Wagner asymptotic analysis **[1, 2]**. With the defined relative radius $\rho = R/R_{cr}$ and the dimensionless time $\tau = \ln(R_{cr})$, Eq. (21) of the main text will transform into its dimensionless expression, yielding as

$$\frac{d\rho}{d\tau} = \nu \frac{(1+\beta\rho)(\rho-1)}{\rho} - \rho, \qquad \text{(A-1)}$$

and the dimensionless expression of continuity equation is

$$P(\rho)\frac{d^2\rho}{d\rho d\tau} + \frac{d\rho}{d\tau}\frac{dP(\rho)}{d\tau} = 3P(\rho) \qquad \text{(A-2)}$$

where $P(\rho)$ is the normalized grain distribution function. At the steady stage, there will be two constraints for the maximum relative radius $\rho_{\max}$, which are

$$\left.\frac{d\rho}{d\tau}\right|_{\rho \to \rho_{\max}} = 0; \quad \left.\frac{d}{d\rho}\frac{d\rho}{d\tau}\right|_{\rho \to \rho_{\max}} = 0, \qquad \text{(A-3)}$$

from which we can calculate the maximum relative radius $\rho_{\max}$, dimensionless constant $\nu$, and the normalized distribution $P(\rho)$ yield as **[3]**

---

\* Corresponding author. zhjwang@nwpu.edu.cn

$$\rho_{\max} = \frac{2}{1-\beta}; \quad \nu = \frac{\alpha M \sigma}{R_{cr}(dR_{cr}/dt)} = \frac{4}{(1+\beta)^2},$$

$$P(\rho) = \frac{Cons \cdot \rho}{[(\beta-1)\rho+2]^{\frac{(5\beta^2+2\beta+5)}{(1-\beta)^2}}} \exp\left(\frac{-6(1+\beta)^2}{(1-\beta)^2[(\beta-1)\rho+2]}\right), \quad \text{(A-4)}$$

The 'kth' moment of the distribution function, $M_k$, is defined as

$$M_k = \frac{\langle R^k \rangle}{R_{cr}^k} = \int_0^{\rho_{\max}} \rho^k \cdot P(\rho) d\rho. \quad \text{(A-5)}$$

And Kamachali et al. **[4]** have defined a single geometrical factor, denoted by $G_f$ here

$$G_f = \frac{\langle R \rangle^2}{\langle R^2 \rangle} = \frac{(M_1)^2}{M_2}, \quad \text{(A-6)}$$

from which the single variable can be ensured, e.g., $\beta = 0.16$ **[3]** with $G_f = 0.87$ **[4]**. Then, we can calculate the thermodynamic constraint of this paper [Eq. (20)], yielding as

$$\int_0^\infty F(R,t) 2\pi R^2 \sigma \left(\frac{\beta \rho}{3} - \frac{\beta \rho^2}{4}\right) dR = 2\pi \sigma \beta R_{cr}^2 \left(\frac{M_3}{3} - \frac{M_4}{4}\right) \approx 0.007 R_{cr}^2, \quad \text{(A-7)}$$

which requires additional term to eliminate it **[5]**, yielding as

$$\int_0^\infty F(R,t) 2\pi R^2 \sigma \left(\frac{\beta_1 \rho}{3} - \frac{\beta_1 \rho^2}{4} + \frac{\beta_2 \rho^2}{4} - \frac{\beta_2 \rho^3}{5}\right) dR$$

$$= 2\pi R^2 \sigma \left[\beta_1 \left(\frac{M_3}{3} - \frac{M_4}{4}\right) + \beta_2 \left(\frac{M_4}{4} - \frac{M_5}{5}\right)\right] = 0 \quad \text{(A-8)}$$

whose corresponding change rate of grain's radius is

$$\frac{d\rho}{d\tau} = \nu \frac{(1+\beta_1 \rho + \beta_2 \rho^2)(\rho-1)}{\rho} - \rho. \quad \text{(A-9)}$$

Now, the new grain size distribution and growth rate can be calculated by the following iterations:

*(i)    Take estimation of Eq. (A-3) as an initial condition, calculate the relationship between $\beta_1$ and $\beta_2$ by Eq. (A-7).*

(ii) With the calculated relation between $\beta_1$ and $\beta_2$, calculate the new $\rho_{\max}$, $\nu$, and $P(\rho)$.

***(It should be mentioned the present solution of $P(\rho)$ is done by numerical solution of Eq. (A-2))***

(iii) Take the refreshed $P(\rho)$ as an initial condition, repeat the former two-step calculations until a set tolerance is reached.

However, it is found the modification of Eqs. (A-8) and (A-9) will break at the second step where there is no solution of the refreshed $\rho_{\max}$ and $\nu$.

As a contrast when we adopt Eqs. (22) and (23) in the main text of the paper, the thermodynamic constraint will become

$$\int_0^\infty F(R,t) 2\pi R^2 \sigma \left( \frac{\beta_1}{3}\rho - \frac{\beta_1}{4}\rho^2 + \frac{2\beta_2}{5}\rho^{\frac{1}{2}} - \frac{2\beta_2}{7}\rho^{\frac{3}{2}} \right) dR$$
$$= 2\pi R^2 \sigma \left[ \beta_1 \left( \frac{M_3}{3} - \frac{M_4}{4} \right) + \beta_2 \left( \frac{2M_{5/2}}{5} - \frac{2M_{7/2}}{5} \right) \right] = 0 \quad \text{(A-10)}$$

and the dimensionless expression of change rate of grain's radius yields as

$$\frac{d\rho}{d\tau} = \nu \frac{\left(1 + \beta_1 \rho + \beta_2 \rho^{\frac{1}{2}}\right)(\rho - 1)}{\rho} - \rho. \quad \text{(A-11)}$$

Applying the former three-step iterations on Eqs. (A-10) and (A-11), we can obtain that the maximum relative radius and dimensionless constant

$$\rho_{\max} = \frac{2}{1-\beta} = 2.56; \quad \nu = \frac{\alpha M \sigma}{R_{cr}(dR_{cr}/dt)} = \frac{4}{(1+\beta)^2} = 3.025, \quad \text{(A-12)}$$

and the first and second moments of the solved grain size distribution in **Fig. 2(a)** of the main text are

$$M_1 = 0.852; \quad M_2 = 0.8382, \quad \text{(A-13)}$$

from which Kamachali et al.'s geometrical factor **[4]** yields as $G_f = 0.8661$, which is the exactly same as the Miyoshi et al.'s ultra-large phase-field simulation results **[6]**. In addition, the growth rate of the squared average radius $K$ is

$$K = \frac{d\langle R \rangle^2}{dt} = (M_1)^2 \frac{dR_{cr}^2}{dt} = 0.4799. \tag{A-14}$$